\documentclass[12pt]{article}

\usepackage{amsmath}
\usepackage[dvips]{epsfig}
\usepackage{amssymb}
\usepackage{color}
\usepackage{colordvi}
\usepackage{graphicx}
\usepackage{pdfsync}

\newtheorem{theorem}{Theorem}[section]

\newtheorem{proposition}[theorem]{Proposition}

\begin{document}

\title{Gravitation and Special Relativity}

\author{D.~H.~Sattinger\\
Department of Mathematics\\
University of Arizona\\
 Tucson, Arizona }
 
 \date{}

\maketitle

 \centerline{\it Dedicated to Klaus Kirchg\"assner}
 
 \bigskip
 
  \centerline{July 5, 2012}
 
 \vskip 1cm
 
 \abstract{ A  mathematical derivation of  Maxwell's equations for gravitation, based on a mathematical proof of Faraday's Law,  is presented.  The theory provides a linear, relativistic Lagrangian field theory of gravity in a weak field, and paves the way to a better understanding of the structure of the energy-momentum tensor in the Einstein Field Equations. Hence it is directly relevant to problems in modern cosmology.  

The derivation, independent of the perturbation theory of Einstein's equations, puts the gravitational and electromagnetic fields on an equal footing for weak fields, contrary to generally held views.  The historical objections to a linear Lorentz invariant field theory of gravity are refuted.}

\vskip 2cm

\noindent davidsattinger@gmail.com \\
\noindent  http://math.arizona.edu/$\thicksim$dsattinger/

\section{Introduction}\label{intro} In 1893 Oliver Heaviside published a paper  \cite{OH} entitled ``A Gravitational and Electromagnetic Analogy''  in which he noted the similarities between the gravitational and electromagnetic fields. ``Now, bearing in mind the successful manner in which Maxwell's localization of electric and magnetic energy in his ether lends itself to theoretical reasoning,'' he wrote,  ``the suggestion is very natural that we should attempt to localize gravitational energy in a similar manner, its density to depend upon the square of the intensity of the force, especially because the law of the inverse squares is involved throughout.'' Heaviside's attempt at a field theory of gravitation was followed by Lorentz (1900) \cite{HAL} and Poincar\'e (1905) \cite{P}.  

Those attempts were abandoned with the success of Einstein's general theory of relativity, which is geometric in nature, and highly nonlinear.  Einstein maintained that gravitational forces were inherently non-linear, and could not be described by a linear, relativistic field theory. The historical arguments against a linear field theory of gravity are discussed in Pais \cite{P} Chapter 13, and in the well-known text {\it Gravitation} by Misner, Thorne, and Wheeler \cite{MTW}, Chapter 7.  

On the other hand, the formal linearization of Einstein's equations at the Minkowski metric gives just such a theory. Chapter 7 of Misner {\it et.al.}, for example,  is entitled ``Incompatibility of Gravity and Special Relativity;" while Chapter 18, is entitled ``The Linearized Theory of Gravity". In \S18.2, p. 442, we find the statement ``The gauge conditions and field equations (18.8a,b) of linearized theory bear a close resemblance to the equations of electromagnetic theory in Lorentz gauge and flat space-time.'' 

As a result of the objections to a linear theory of gravitation, the early theories of gravity have been reincarnated under the rubric ``Gravito-Electromagnetic Analogy,'' in which linear field theories of gravity are obtained as a formal perturbation of the Einstein Field Equations in the vicinity of the Minkowski metric. But what is the physical significance of such formal perturbation methods if no  {\it bona fide} linear relativistic field theory of gravitation exists? The perturbative approach is therefore a tacit acknowledgment that a linear field theory of gravity for weak fields must exist; and indeed it does.  We shall call it the Maxwell-Heaviside theory, and shall show that the equations are not only mathematically rigorous \S\ref{gravitation}, but utilitarian as well \S\ref{galaxies}; and moreover, that Einstein's objections to them do not apply to weak fields \S\ref{enigma}.

 ``It is amusing to recall,'' write  Clark and Tucker \cite{CT}, ``that one of the first theories of post-Newtonian gravitation was formulated by Heaviside in direct analogy with the  theory of electromagnetism.  $\dots$ It predicted that gravitation, like electromagnetism, was mediated by an independent vector field rather than with a second-degree tensor field associated with the metric of space-time. This difference [implies] that the analogy between weak gravity and electromagnetism is incomplete.'' 
 
Maxwell's derivation of his equations for the electromagnetic field was based on extensive empirical data,  especially Faraday's Law of electromagnetic induction -- the mathematical cornerstone of the theory. In the case of gravitation, however, there is as yet no experimental evidence for a gravitational field induced by the motion of mass; but a mathematical proof of Faraday's Law is given in \S\ref{gravitation}, providing the starting point for a derivation of Maxwell's equations for gravity. This puts the gravitational and electromagnetic fields on an equal footing for weak fields, thus completing the Heaviside analogy. 

One should expect that the linearized Einstein equations coincide with the Maxwell-Heaviside theory, but the matter is not straightforward, as Clark and Tucker demonstrate:  ``The question of the gauge transformations of the perturbative Einstein equations $\dots$ leads one to contemplate the {\it most useful way} to define the {\it gravito-electromagnetic} fields in terms of the perturbed components of the space-time metric. Different choices are often responsible for the location of odd factors of four that permeate the {\it gravito-electromagnetic} equations compared with Maxwell's equations.''

Einstein's General Theory of Relativity models the gravitational field as the geodesic flow of a metric tensor $g^{\mu\nu}$, coupled to the energy-momentum tensor $T^{\mu\nu}$ by a highly nonlinear set of partial differential equations known as the Einstein Field Equations:
\begin{equation}\label{EFE}
R^{\mu\nu}-\frac12 R\, g^{\mu\nu}=\frac{8\pi G}{c^4}T^{\mu\nu}.
\end{equation}
Here $R^{\mu\nu}$ is the Ricci curvature tensor of $g^{\mu\nu}$, $R$ is the scalar Ricci curvature,  and $G$ is the gravitational constant and $c$ is the speed of light. Roughly speaking,  $g^{\mu\nu}$ describes the geometry of space-time,  $T^{\mu\nu}$ the physics. 

Einstein originally obtained a solution of his equations for a point mass, for which $T^{\mu\nu}=0$ everywhere except the origin.  He obtained correct results for the advance of the perihelion of Mercury, and the deflection of light by the Sun. His work was followed almost immediately by that of Schwarzschild \cite{schw}, who obtained the metric tensor for Einstein's solution. There are now many other examples of solutions of Einstein's equations in a vacuum. These include the Kerr black hole, the Neugebauer-Meinel disk, and more generally, a system of integrable equations known as the Ernst equation which includes these two, Lenells \cite{JL}. 

Since roughly 99\% of the mass in the solar system lies in the Sun, the Schwarzschild metric is an excellent model for the dynamics of the Solar System as motion in a central force field. The Schwarzschild metric has proved to be one of the most remarkable models of Mathematical Physics; but, due to the distribution of mass and energy in a galaxy, it fails to model the dynamics of  galaxies. For this reason, the energy-momentum tensor plays a fundamental role in galactic dynamics, including the questions of dark matter and dark energy (see \S \ref{galaxies}). It is also crucial to the study of  gravitational collapse, beginning with the celebrated papers of Tolman \cite{Tolman} and Oppenheimer and Synder \cite{Op}. Yet the energy-momentum tensor is known explicitly only in the case of electrodynamics: it is the Maxwell stress tensor. The fact that gravitation can also be modeled by Maxwell's  equations  provides significant additional information as to the structure of  $T^{\mu\nu}$.  

Maxwell's equations constitute potential theory in four dimensional space-time.  Differential forms and the Hodge star operation are the natural language of potential theory, and extensive use will be made of them here. There are two orientations of an orientable manifold, and consequently two Hodge star operations; the orientation of space-time is reversed in going from the electromagnetic to the gravitational field. With a view to making this paper self-contained, a brief operational introduction to differential forms and the Hodge star operation is given in \S \ref{forms}.  A fuller account is given in  \cite{MTW}; and an exposition of the exterior differential calculus along the lines used here is given in Sattinger and Weaver \cite{SW}. A review of Maxwell's equations for electrodynamics in the language of differential forms is given in \S\ref{electro}.

Minkowski\footnote{Minkowski died suddenly in 1909. The paper cited here is a posthumous publication of the article that appeared in the {\it G\"ottinger Mathematischen Gesellschaft} in 1907.} \cite{Mink} did not use a metric tensor in his famous 1907 construction of space-time. Instead he represented it as $\mathbb E^4$ with $x^4=ict$; Stratton \cite{JAS} does the same. It will be denoted here by $\mathbb M^4$.  Minkowski's complex structure permits the use of the Hodge star operation associated with $\mathbb E^4$ rather than that tied to the Minkowski metric, and simplifies the presentation.  The complex structure is inherited by Maxwell's equations; and the Lorentz group is obtained from the rotation group on $\mathbb E^4$ under the transformation $x_4\to ict$.

\section{Mathematical Preliminaries}\label{forms}

We assume the reader is already familiar with the basic operations of wedge product $\wedge$ and exterior derivative $d$ on $p$  forms $\Lambda_p$ and that the theorems of Green, Gauss, and Stokes are collected in a single theorem, known as Stokes' theorem
\begin{equation}\label{stokes} \iint\limits_\Omega d\omega=\int\limits_{\partial \Omega} \omega.
\end{equation} Here, $\omega\in \Lambda_p$ has differentiable
coefficients, and $\Omega$ is a $p+1$ dimensional, oriented manifold embedded in
$\mathbb E^n$, with smooth boundary $\partial \Omega$. 

A  form $\omega\in \Lambda_p$ is said to be {\it closed} if $d\omega=0$ and {\it exact} if $\omega=d \chi$, where $\chi\in\Lambda_{p-1}$. Since $d^2=0$, a $p$-form is closed if it is exact. In a simply connected region the two conditions are equivalent; and it will be sufficient to restrict ourselves to this case. A necessary and sufficient condition for $\omega\in \Lambda_p$ to be exact in a region $U$ (not necessarily simply connected) is that its integral over every closed $p$ manifold $\Sigma\subset U$ vanish:
$$
\iint\limits_{\Sigma}\omega =0, \quad \text{whenever} \quad \partial\Sigma=\emptyset.
$$

In the special case of a 1-form $E$, $\Sigma$ is a closed path, and the integral above is a line integral called the {\it circulation}. If the circulation vanishes for every smooth closed path, then regardless of the topology of the region, $E$ is exact, and there exists a 0-form $\phi$ (that is, a single-valued function) such that $E=-d\phi$. It is standard convention to normalize the potential to vanish at infinity, so that it is  explicitly given by
\begin{equation}\label{potenergy}
\phi({\bf x})=\int_{\bf x}^\infty E, \qquad {\bf x}\in \mathbb E^3.
\end{equation}
The electrostatic and gravitational fields are both conservative, so that their corresponding 1-forms  are exact; but the electrostatic potential is positive, while the gravitational potential is negative. This reflects the fact that the gravitational force is attractive, while the electrostatic force is defined in terms of  like charges and so is repulsive.

The Hodge star operation $\ast$ on differential forms over an $n$ dimensional orientable manifold $\mathcal M$ plays a fundamental role in potential theory. It is defined as follows: Given an oriented volume element $dv$ on $\mathcal M$ and  $\omega\in \Lambda_p$, $\ast \omega$ is defined as the $n-p$ form for which $\omega\wedge \ast \omega=dv$. 
The standard (right-handed) volume element on $\mathbb E^3$ is $dv=dx^1\wedge dx^2\wedge dx^3$. The associated Hodge star operation is 
\begin{equation}\label{*3}
\ast dx^i=dx^j\wedge dx^k, \qquad \ast 1=dv, \qquad \ast\ast=id.
\end{equation}
($i,j,k$ in cyclic order.) 
We shall always denote by $\ast$ the Hodge star operation on $\mathbb E^n$ associated with the standard  volume element $dv=dx^1\wedge \dots \wedge dx^n$.

There is a second volume element on $\mathbb E^3$, namely the left-handed volume element $d\tilde v=dx^3\wedge dx^2\wedge dx^1$, and hence a second star operation, denoted by $\tilde\ast$.  Every orientable manifold has two orientations, corresponding to the even and odd permutations of the basis 1-forms, and consequently two star operations associated with it. Since 
$ \xi\wedge \widetilde\ast\,\eta=-\xi\wedge\ast\,\eta$
for all basis form $\xi,\eta\in \Lambda_p$, it follows that $\tilde\ast=-\ast$. The two orientations of $\mathbb E^4$ play a fundamental role in distinguishing the gravitational and electromagnetic fields. 

For any star operation $\ast$, an inner product, called the Hodge duality, is defined for $\xi,\,\eta\in \Lambda_p(\mathbb E^n)$ by
\begin{equation}\label{Hdual}
(\xi,\eta)=\iiint\limits_{\mathbb E^n} \xi\wedge \ast \eta, \qquad \xi,\eta\in \Lambda_p
\end{equation}
The Hodge duality in turn defines a formal adjoint to the exterior derivative $d$.  Called the {\it coderivative}, it maps $\Lambda_{p+1}$ to $\Lambda_p$, and is defined implicitly by the relation 
\begin{equation}\label{adjoint}
(d\,\xi,\eta)=(\xi, \delta\eta) \qquad  \xi\in \Lambda_p,\ \eta\in\Lambda_{p+1}
 \end{equation} 
where $\xi,\,\eta$ have compact support on $\mathbb E^n$.

Unless otherwise stated, we shall denote the coderivative associated with the standard Hodge star operation $\ast$ by $\delta$ and that associated with the operation $\tilde\ast$ by $\tilde\delta$. The action of $\delta$ on $\Lambda_p$ will be denoted by $\delta_p$.

\begin{proposition}\label{coder} We have 

 i) $\delta_p=(-1)^{p-1} \ast^{-1} d\,\ast$ on $\mathbb E^n$ for both $\ast$ operators;
 \smallskip
 
 ii) $\tilde \ast=-\ast$ on $\mathbb E^n$, hence $\tilde\delta_p=\delta_p$;
 
 \smallskip
  % ii) $\tilde\ast=-\ast$; hence $\delta=\tilde\delta$ on $\mathbb E^n$; 
 iii)   $\delta_p=(-1)^{p-1}\ast\,d\,\ast$  on $\mathbb E^3$;  
 
  \smallskip
 iv)  $\ast\,\ast =(-1)^p\,id$ and $\delta=\ast\,d\,\ast$ on $\mathbb E^4$.
\end{proposition}

\noindent{\sl Proof:}  i):  Let $\xi,\,\eta$ be smooth differential forms with compact support, and integrate the identity
\begin{multline*}
d(\xi\wedge\ast \eta)=d\xi\wedge \ast \eta+(-1)^p\xi\wedge d\ast\eta\\[4mm]
=d\xi\wedge\ast\eta-\xi\wedge \ast (-1)^{p-1} \ast^{-1} d\ast\eta
=d\xi\wedge\ast\eta-\xi\wedge\ast\delta\eta
\end{multline*}
over any large ball $B$. By Stokes' theorem, we get on the left side
$$
\iiint\limits_B d(\xi\wedge\ast\eta)=\iint\limits_{\partial B} \xi\wedge\ast\eta=0
$$
when $B$ is sufficiently large. The right side then gives \eqref{adjoint}. The same proof holds for $\tilde\ast$.
\smallskip

ii):   The assertion for $1\le p\le n-1$  follows from the fact that  $\xi\, \wedge\,\ast\,\eta=-\xi\wedge\,\tilde\ast\,\eta$ for all $\xi,\eta\in \Lambda_p$ for such $p$.  For $p=0,n$, note that $\tilde \ast 1= d\tilde v=-dv$ while $\tilde\ast d\tilde v=1$ is equivalent to
$$
(d\tilde v,d\tilde v)_U= \int_U d\tilde v \wedge \tilde \ast d\tilde v=\int_U d\tilde v=-\int_Udv=-{\rm vol}\,( U).
$$
iii): From \eqref{*3} it follows that $\ast^{-1}=\ast$ on $\mathbb E^3$, and the assertion in iii) follows from   that in i). 

\smallskip

iv): The Hodge star operation on $\mathbb E^4$ associated with the standard volume element is
\begin{alignat}{2}
&\ast dx^j= dx^k\wedge dx^l\wedge dx^4 \qquad &\ast\, dx^4=-dv \label{1form}\\[4mm]
&\ast dx^j\wedge dx^k= dx^l \wedge dx^4  &\ast \,  dx^j\wedge  dx^4=dx^k\wedge dx^l  \label{2form}\\[4mm]
&\ast dv=dx^4,  &\ast\,  dx^j\wedge dx^k\wedge dx^4=-dx^l.\label{3form}
\end{alignat}
It follows that $\ast^{-1}=(-1)^p\, \ast$ on $\Lambda_p(\mathbb E^4)$. On the other hand,
$$
\ast\,:\, \Lambda_p\to \Lambda_{n-p}, \quad d\,: \Lambda_{n-p}\to\Lambda_{n-p+1}, \quad 
\ast^{-1}\,: \Lambda_{n-p+1} \to \Lambda_{p-1}.
$$
For $n=4$, therefore, $(-1)^{p-1}\ast^{-1}=\ast$; and so, by i),
\begin{equation}\label{deltaE4}
\delta_p=(-1)^{p-1}\ast^{-1}d\,\ast =\ast\,d\,\ast \ \text{on}\ \mathbb E^4\ \text {for all}\ p 
\end{equation}
This completes the proof of Proposition \ref{coder}.
\medskip

\medskip

Maxwell's equations are formulated as a system of partial differential equations for a set of ``vector'' fields ${\bf E}, {\bf D},{\bf H},{\bf B}$; but the vector notation is ambiguous. The electric and magnetic fields,  ${\bf E}$ and ${\bf H}$, are ordinary polar vectors, and are identified with the 1-forms $E={\bf E \cdot dx},\  H={\bf H \cdot dx}$.

The ``vectors'' ${\bf B}$ and ${\bf D}$, the magnetic induction and electric displacement, are axial vectors, associated with surface integrals (fluxes), hence with 2-forms $B,D$. In $\mathbb E^3$, we take $B= {\bf B\cdot d S}$ and similarly for $D$.  The expression $d{\bf S}$ denotes the vector element of surface area on a 2 dimensional surface $S$ embedded in $\mathbb E^3$; it is given by $d{\bf S}=(X_u\times X_v)\, du\wedge dv$, where $X=(x^1(u,v),x^2(u,v),x^3(u,v))$ is a parametrization of a neighborhood of $S$ by local coordinates $u,v$. We leave it to the reader to verify the identities
$$
dx^i\wedge dx^j=\frac{\partial (x^i,x^j)}{\partial (u,v)} du\wedge dv,
\qquad B=B_j\,dx^k\wedge dx^l.
$$
Here and throughout this article, the expression for $B$ signifies a summation over $j,k,l$ from 1 to 3 in cyclical order.

\section{Potential theory in $\mathbb E^3$}
There are two physical parameters in Maxwell's theory, the electric and magnetic inductive capacities $\epsilon$ and $\mu$. Stratton defines them by the constitutive relations
\begin{equation}\label{CR}
{\bf D}=\epsilon {\bf E}, \qquad {\bf B}=\mu {\bf H},
\end{equation}
relating the axial vectors ${\bf D}$ and ${\bf B}$ to the polar vectors ${\bf E,\  H}$. This logical inconsistency is removed by defining $\epsilon$ and $\mu$ in terms of the associated differential forms. Accordingly,  Gauss'  law of electrostatics in an isotropic medium is stated in terms of the Hodge star operation as:
\begin{equation}\label{GaussE}
\iint\limits_S D=Q_S, \qquad D=\epsilon \ast E .
\end{equation}
Here, $Q_S$ is the total charge contained inside $S$,  $E$ is the 1-form $E={\bf E}\cdot d{\bf x},$ where ${\bf E}$ is the electric field, and  $D$ is Maxwell's displacement. 

For a point charge $q$ at the origin
\begin{equation}\label{E,D}
E=\frac{G_e\,q }{r^2}\,{\bf \hat r}\cdot d{\bf x}=\frac{G_e\,q }{r^3}x_j dx^j; \qquad D=\frac{q\, x_jdx^k\wedge dx^l}{4\pi r^3}.
\end{equation}
where $G_e$ is a physical constant, determined by Coulomb. The first equation is Coulomb's inverse square law. The integral of $D$ over a closed surface $S$ enclosing the origin can be deformed to an integral over a sphere of radius $R$ centered at the origin. By Stokes' theorem
$$
\iint\limits_{S} D=\frac{q}{4\pi R^3}\iint\limits_{S} x_jdx^k\wedge dx^k=
\frac{q}{4\pi R^3}\iiint\limits_{B_R} 3\, dv =q,
$$
where $B_R$ denotes the interior of the sphere. The second equation in  \eqref{GaussE}, together with the expressions for $D$ and $E$ in \eqref{E,D}  imply  $\epsilon=1/4\pi G_e$. 

In the case of a continuous charge density, we have 
\begin{equation}\label{divD}
D=D_j dx^k\wedge dx^l, \quad dD=\rho\,dv, \quad \rho=\sum_j \frac{\partial D_j}{\partial x^j}.
\end{equation}
The differential expression for $\rho$ is valid in general coordinates, but in Cartesian coordinates it can be written in vector form ${\rm div} \,{\bf D}=\rho$. If $Q_S>0$ throughout a region $U$, then $\rho\ge 0$ in $U$.

The equations of the electrostatic field are $E=-d\phi,\ D=\epsilon\ast\,E.$ Eliminating $D$ using Proposition \ref{coder}, iii) we find $\delta E=\rho/\epsilon.$ Combining these two equations for $E$ we obtain $\delta\,d\,\phi=-\rho\slash\epsilon.$
It is easily verified that on $\Lambda_0(\mathbb E^3)$ (functions)
$$
\delta\, d = \ast \,d\ast\,d =\sum_{j=1}^3\frac{\partial ^2 }{\partial x_j^2}, 
$$
and the equation for the electrostatic potential on $\mathbb E^3$ is $\Delta\phi=-\rho \slash\epsilon.$

\medskip

In the case of gravitation, the lines of force flow into the region bounded by $S$, since the force is attractive. If we follow the procedure above, \eqref{divD} will lead to $\rho=\varepsilon \,{\rm div}\,{\bf F}<0$. A positive mass density can be obtained by reversing the orientation, and using  $\tilde \ast$. It is given by
$$
\tilde \ast dx^i=dx^j \wedge dx^k,  \ \  i,j,k\  \text{anti-cyclic}; \qquad \tilde \ast 1= d\tilde v; \qquad \tilde \ast
\tilde\ast =id.
$$

Gauss' law for the gravitational field then takes the form \begin{equation}\label{GaussG}
\iint\limits_S D=M,  \qquad D=\epsilon_g\, \tilde \ast\, F, \qquad \epsilon_g=\frac1{4\pi G}
\end{equation}
where $F$ is the 1-form associated with the gravitational field, $M$  the total mass in $S$,  and $G$ the Cavendish gravitational constant. The mass density $\rho$ is non-negative, due to our choice of the left-handed volume element, for now
$$
dD=\varepsilon( \,{\rm div}\, {\bf F}) \,d\tilde v=\rho\,dv, \qquad \rho=-\varepsilon\,{\rm div} {\bf F}>0.
$$

%$D=\varepsilon\,\tilde \ast \, F$, where $D$ is the gravitational displacement, $F={\bf F}\cdot d{\bf x}$, and $\varepsilon$ is the ``gravitational permittivity.''  It is easily verified that
%We thus take  the 2-form associated with the gravitational displacement to be  $D=D_i\,dx^j\wedge dx^k$, where $i,j,k$ are now in anti cyclic order. We then get  Equation \eqref{divD} now becomes
%$$dD=\sum_j \frac{\partial D_j}{\partial x^j} dx^l\wedge dx^k}=\div {\bf D} \,d\tilde v=

Summarizing the two cases, we have
\begin{theorem}\label{pot}  For the electrostatic field in $\mathbb E^3$, take the Hodge star operation associated with the right-handed volume element. The equations for the electrostatic potential are then
\begin{equation}
E=-d\phi, \qquad \delta E={\rm div}\, {\bf E}=\frac\rho\epsilon, \qquad \Delta \,\phi=-\frac\rho\epsilon. \label{poissonE}
\end{equation}

For the gravitational field take the Hodge star operation $\tilde \ast$ associated with the left-handed volume element. The equations for the gravitational potential  are then
\begin{equation}\label{poissonG}
F=-d\phi, \qquad \tilde\delta F={\rm div}\, {\bf F}=-\frac\rho{\epsilon_g}\, \qquad \widetilde\Delta \phi=\Delta\phi=\frac\rho{\epsilon_g}.
\end{equation}
\end{theorem}
The proof of the gravitational case is left to the reader.

\section{Potential Theory in $\mathbb M^4$, I}\label{electro}
Maxwell's equations of electrodynamics in vector form are \cite{JAS} 
\begin{alignat}{2}
& {\bf \nabla \times E}+\frac{\partial {\bf B}}{\partial t}=0, \qquad & {\rm div} \,{\bf B}=0; \label{Faraday}
\\[4mm]
&  {\bf \nabla \times H}-\frac{\partial {\bf D}}{\partial t}={\bf J},  & {\rm div}\, {\bf D}=\rho.
  \label{Ampere}
 \end{alignat}
 The system is closed with the two constitutive relations \eqref{CR}.  The first pair of equations is the differential form of Faraday's law of magnetic induction, the second Amp\`ere's law modified by Maxwell's introduction of the displacement current ${\bf D}_t$;  $\rho$ denotes the charge density, and ${\bf J}=\rho{\bf v}$ the current.
Classical potential theory begins with a 1-form $F$ with zero circulation;  the corresponding object in Maxwell's equations is the Faraday 2-form 
\begin{equation}\label{Farint}
F=E\wedge dt + B,
\end{equation} 
where, as above, $ E={\bf E}\cdot d{\bf x} $ and $B={\bf B} \cdot d{\bf S}$.

\begin{theorem} Assume $\bf E$ and $\bf B$ are smooth vector fields. Then Faraday's Law \eqref{Faraday} is a necessary and sufficient condition that the Faraday 2-form \eqref{Farint} be exact. In that case there is a 1-form $A=A_jdx^j$ such that $F=dA$. \end{theorem}
\noindent{\sl Proof:} We calculate $dF=dE\wedge dt+dB$ in Cartesian coordinates:
\begin{align*}
dE=&\frac{\partial E_i}{\partial t}  dt \wedge dx^i +\left(\frac{\partial E_i}{\partial x_j}-\frac{\partial E_j}{\partial x_i}\right)dx^j\wedge dx^i\\[4mm]
=&(\nabla\times {\bf E})_i \,dx^j\wedge dx^k-\frac{\partial E_i}{\partial t}\,dx^i\wedge dt ;  \\[4mm]
dB=&\frac{\partial B_i}{\partial t} dx^j\wedge dx^k\wedge dt+\frac{\partial B_i}{\partial x^i} dx^i\wedge dx^j\wedge dx^k\\[4mm]
=&\frac{\partial B_i}{\partial t} dx^j\wedge dx^k\wedge dt+\text{div}\,{\bf B}\,dv
\end{align*}
Thus $dF=0$ implies the two equations in \eqref{Faraday}. 

Conversely, Faraday's Law implies that $F$ is closed; and since $\mathbb M^4$ is simply connected, a  closed form is exact. In short, Faraday's Law is equivalent to the existence of a 1-form $A$ such that $F=dA$. $\blacksquare$

\smallskip

We shall call the class of all 1-forms $A=A_\mu\,dx^\mu$ for which $\partial_\mu\,A_\mu=0$ Lorentzian 4-potentials, and we denote them by $\mathcal P$.  Throughout this paper we consider only such 4-potentials; and  we restrict the discussion to smooth forms and vector fields.  

The 4-current $J$ is defined to be $J=J_j dx^j={\bf J}\cdot d{\bf x} +J_4dx^4$.  Under the transformation $x_4=ict, \ J_4=ic\rho$,  $J=\rho v_jdx^j-c^2\rho\, dt$ (see Stratton, \S1.21) The conservation of charge in rectangular  coordinates is ${\rm div}\, {\bf J}+\partial_t \rho=0$. This can be written as $\partial_\mu J_\mu=0$; hence the 4-current also belongs to the class $\mathcal P$.

By \eqref{2form} the pair of constitutive relations \eqref{CR} can be written as
\begin{equation}\label{constitutive_Max} 
D=\epsilon \ast\,E\wedge dx^4, \qquad B=\mu  \ast H\wedge dx^4.
\end{equation}

 \begin{theorem}\label{Maxforms} The Maxwell-Amp\`ere law is equivalent to the equation $ d G=\ast J$, where $G= ic\,(H\wedge dt-D)$ and $J$ is the 4-current. The necessary and sufficient condition for the solvability of this equation is $d \ast J=0$.

Assume the Faraday 2-form is exact, and that the constitutive laws  \eqref{constitutive_Max} hold.  Then $\ast F=\mu G$ if and only if  $\epsilon\mu c^2=1;$ and in that case Maxwell's equations take the form  
\begin{equation}\label{emequations}
F=dA, \qquad \delta F=-\mu J, \qquad \square A=\mu J, 
\end{equation}
where
\begin{equation}\label{dalembertian}
\delta d A=\ast\, d\ast d A=-\square\, A , \qquad \square = \Delta -\frac1{c^2}\frac{\partial^2}{\partial t^2}.
\end{equation}
\end{theorem}

\noindent {\sl Proof:} Observe that  $G$, with $H=H_jdx^j$ and $D=D_j\,dx^k\wedge dx^l$, represents a general 2-form on $\mathbb M^4$. By direct calculation
\begin{equation}
dG= \left(\frac{\partial H_l}{\partial x^k}-\frac{\partial H_k}{\partial x^l}-ic \frac{\partial D_j}{\partial x^4}\right)dx^k\wedge dx^l\wedge dx^4-ic\sum_{j=1}^3 \frac{\partial D_j}{\partial x^j}\,dv   \label{dG}
\end{equation}
In Cartesian coordinates, 
\begin{equation}\label{dG*J}
\frac{\partial H_l}{\partial x^k}-\frac{\partial H_k}{\partial x^l}=\left({\bf \nabla \times H}\right)_j, \qquad \sum_{j=1}^3 \frac{\partial D_j}{\partial x^j}={\bf {\rm div}\, D}.
\end{equation}
By \eqref{1form} $\ast J= (J_j\,dx^k\wedge dx^l\wedge dx^4 -ic\rho\,dv)$;  hence in Cartesian coordinates \eqref{dG} is precisely the Maxwell-Amp\`ere equation \eqref{Ampere}. Note that the conservation of charge implies that $d\ast\,J=0.$

Now
\begin{align*}\ast\,F=&\ast \left(\frac{E}{ic}\wedge dx^4+B\right)=\frac1{ic\epsilon}D
+\mu H\wedge dx^4\\[4mm]=&\mu\left(H\wedge dx^4+\frac1{ic\epsilon\mu}D\right) =\mu\,ic\left(H\wedge dt+\frac1{(ic)^2\epsilon\mu}D\right)=\mu\,G,   
 \end{align*}
if and only if $\mu\epsilon c^2=1$. By Proposition \ref{coder}, iv):
 $$ 
\delta F=(\ast\,d\,\ast)\ast\,\mu G = \mu\ast dG=\mu \ast\ast\,J=-\mu J.
$$

The third equation in \eqref{emequations} follows from the relation $\delta\,d=\ast d\ast d=-\square.   \ \blacksquare$

\section{Potential Theory in $\mathbb M^4$, II}\label{gravitation}

In the case of gravitation there is presently no experimental analog of Faraday's Law of magnetic induction; but it turns out that Faraday's Law is a mathematical artifact of the geometry of space-time.  Though the following is general, we retain the notation of electrodynamics for clarity. 
\begin{theorem}\label{Max_geometry} Let $E=E_j(x,t)dx^j$ be any space-like 1-form on $\mathbb M^4$. Then there is a family of Lorentzian 4-potentials  $\mathcal P$ and an exact 2-form $B=B_j(x,t)\,dx^k\wedge dx^l$  such that $F=E\wedge dt+B=dA$ for all $A\in\mathcal P$.
 \end{theorem}

 \noindent{\sl Proof:}  The 4-potentials $A \in \mathcal P$ are obtained by inverting the linear hyperbolic system of partial differential equations
\begin{equation}
\begin{pmatrix} \partial_t &0&0& \partial_1\\[2mm]
0& \partial_t  &0& \partial_2\\[2mm]
0&0& \partial_t  & \partial_3\\[2mm]
  \partial_1 &  \partial_2&  \partial_3&c^{-2}\partial_t \end{pmatrix}
     \begin{pmatrix} A_1\\[2mm] A_2 \\[2mm]A_3\\[2mm] \phi \end{pmatrix}
   +\begin{pmatrix} E_1\\[2mm] E_2 \\[2mm]  E_3\\[2mm] 0 \end{pmatrix}=0, \label{4potential}
       \end{equation}
where $ \partial_j=\partial/\partial x^j$ and $\phi=-icA_4.$

 A quick calculation shows that for  $A=A_jdx^j$,
   \begin{align*}
   dA=&
   \sum_{j<k\le 3}\left(\frac{\partial A_k}{\partial x^j}-\frac{\partial A_j}{\partial x^k}\right) dx^j\wedge dx^k\\
 &  \hskip.75in +\sum_{j=1}^3 \left(\frac{\partial A_4}{\partial x^j}-\frac{\partial A_j}{\partial x^4}\right) dx^j\wedge dx^4.
   \end{align*}
Putting $F=dA$ we obtain
  \begin{equation}\label{BE}
    B_i=\frac{\partial A_k}{\partial x^j}-\frac{\partial A_j}{\partial x^k},
  \qquad
  \frac{E_j}{ic} =\frac{\partial A_4}{\partial x^j}-\frac{\partial A_j}{\partial x^4}. 
  \end{equation}
  The first set of equations implies that $\partial_iB_i=0$ no matter the choice of  $A$. Adding the Lorentz condition $\partial_\mu A_\mu=0$ and  putting   $\phi(x,t)=-icA_4$,  we arrive at \eqref{4potential}. 
  
The system \eqref{4potential} constitutes a first order hyperbolic system. Eliminating the $A_j$, we obtain
  \begin{equation}\label{wave}
  \Delta \phi-\frac1{c^2}\phi_{tt}+\rho=0,  \qquad \rho=\frac{\partial E_i}{\partial x_i}, \qquad \Delta\phi=\sum_{j=1}^3\partial_j^2\phi.
  \end{equation}
Given a solution $\phi$, set $\psi(x,t)=\int^t \phi(x,s)ds$. Then
\begin{equation}\label{Aj}
A_j=-\int^t E_j\,ds -\frac{\partial \psi}{\partial x^j}, \ \ j=1,2,3; \quad A_4=-\frac{\partial \psi}{\partial x^4}.
\end{equation}

Any two solutions of \eqref{4potential} differ by a homogeneous solution -- i.e. $E=0$, hence $\rho=0$ and $\phi$ is a  solution of the homogeneous wave equation. By \eqref{Aj}, the 4-potential of a homogeneous solution is then$-d\psi$. But $dA$ is unchanged under the gauge transformation $A\to A-d\psi$; hence $B$ does not depend on the choice of solution. $\blacksquare$
\medskip

Remark: The 2-form $B$, though independent of the choice of $A\in\mathcal P$, is not uniquely determined. Consider a divergence-free, stationary vector field ${\bf V}=(V^1,V^2,V^3)$, and put $V=V_jdx^j$. Then $V$  is a  Lorentzian 4-potential and under the gauge transformation $A\to A+V$, $B$ goes to $B+B_V$, where $B_V$ is obtained from $V$ via \eqref{BE}. In the electromagnetic case, $B_V$ is a magnetostatic field, generated by a steady current, while $B$ corresponds to the induced field, generated by fluctuations in the electric field. 

\medskip

The inductive capacity $\epsilon_g$ for gravity was given in \eqref{GaussG}.  The recognition that the mass current, analogous to the charge current in electrodynamics, is the momentum, is due to Heaviside \cite{OH}.  The  momentum 1-form for a particle in Special Relativity is (\cite{LL} \S9):  
\begin{equation}\label{4mom}
P=p_\mu dx^\mu={\bf p}\cdot d{\bf x}+ic\rho\,dx^4=\rho v_j dx^j +ic\rho dx^4,
\end{equation}
where $\rho=\rho_0 (1-v^2/c^2)^{-1/2}$ with $\rho_0$  the rest mass density.  

As in the case of static potential theory on $\mathbb E^3$, we need to reverse orientations in going from the electromagnetic to the gravitational field. For gravitation we use the Hodge star operator associated  with the oriented volume element $dx^4\wedge dv$:
\begin{alignat}{2}
&\widetilde\ast\, dx^j= -dx^4\wedge\, dx^k\wedge dx^l  \qquad &\widetilde\ast \,dx^4=dv \label{1form-}\\[4mm]
&\widetilde\ast \,dx^j\wedge dx^k= dx^4 \wedge dx^l  &\widetilde\ast \,  dx^4\wedge  dx^j=dx^k\wedge dx^l  \label{2form-}\\[4mm]
&\widetilde\ast\, dv=-dx^4,  &\widetilde\ast\,  dx^4\wedge dx^j\wedge dx^k=dx^l \label{3form-}.
\end{alignat}

We use the same 2-forms $F$ and $G$ that we used in the electromagnetic case, but replace $\ast$ by $\tilde\ast$ in the computations. This time we have to solve 
$$
dG=\tilde\ast P, \qquad \tilde\ast P=-p_j\,dx^4\wedge dx^k\wedge dx^l+ic\rho\,dv.
$$ 
 The vector form of  the Maxwell-Heaviside equation -- the analog of the Maxwell-Amp\`ere equation for the gravitational field -- is then
 \begin{equation}\label{MaxHeaviside}
  {\bf \nabla \times H}-\frac{\partial {\bf D}}{\partial t}=-{\bf p}, \qquad {\rm div}\, {\bf D}=-\rho.
 \end{equation}
 
By Proposition \ref{coder} ii) we have $\ast\tilde F=-\ast\,F=-\mu G$.; but we may also check this by a direct computation using the equations for $\tilde\ast$ above:
 \begin{align*}
 \tilde \ast F=&\tilde\ast \left[ \frac{E_j}{ic} dx^j\wedge dx^4+B_j dx^k\wedge dx^l\right] \\[4mm]
 =&- \left[ \frac{D_j}{ic\epsilon_g} dx^k\wedge dx^l+ic\mu H_j dt\wedge dx^j\right] \\[4mm]
 =& -ic\mu\left[ H\wedge dt+\frac{D_j}{(ic)^2)\mu_g\epsilon_g} dx^k\wedge dx^l \right]
 =-\mu G.
 \end{align*}
  Therefore  $ \tilde \delta F=\tilde\ast d\,\tilde\ast F=-\mu\tilde\ast\,dG=-\mu \tilde\ast \tilde\ast P=\mu P.$
 The other equations follow as before. 
 
The Maxwell-Heaviside equations are given in the following:
\begin{theorem}\label{MaxGr} Let $F$ and $G$ be the 2-forms in Theorem \ref{Maxforms}
and let $P$ the mass current \eqref{4mom}. The necessary and sufficient condition for the solvability of the equation $dG=\tilde\ast P$  is $\tilde\delta P=0$.  Let  $\mu_g=4\pi G c^{-2}$. Then the Maxwell-Heaviside equations for gravity are 
\begin{equation}\label{MHtilde}
F=dA, \qquad \tilde\delta F=\mu_g P\, \qquad \widetilde\square A=-\mu_g P.
\end{equation}
In terms of the $\ast$ operation, the equations are
\begin{equation}\label{MH}
F=dA, \qquad \delta F=\mu_g P\, \qquad \square A=-\mu_g P,
\end{equation}
 \end{theorem}
   \medskip
   
Remarks: i) The hypothesis $\mu_g=4\pi G c^{-2}$ is equivalent to  the assumption that gravitational waves travel with the speed of light. This relationship was known experimentally in the theory of the electromagnetic field and was cited explicitly by Maxwell \cite{Maxwell} (pp. 577-580). In the theory of Special Relativity, $c$ appears as a universal constant of nature that determines not only the speed of light, but the speed of gravitational waves as well in a vacuum. In Einstein's general theory of relativity, it comes out of the analysis of weak gravitational waves \cite{MTW}. 

\smallskip

ii) The value of $\mu_g$ is
$$
\frac{4\pi G}{c^2}=9.31\times 10^{-28}\frac{ cm\, sec^2}{ gm}
$$
The extreme smallness of this parameter indicates why the gravitational field induced by the motion of mass is so difficult to detect.

\section{The Lagrangian}\label{Lagrangian}

Theorems \ref{Maxforms} and \ref{MaxGr} can be merged into a single statement.
\begin{theorem}\label{E&G}
Let $P$ be the 4-momentum \eqref{4mom}, $e$ be the electric charge, and $\ast$ the Hodge star  corresponding to $dv\wedge dx^4$. Then Maxwell's equations are
\begin{gather}
J=(-e)^\sigma P, \qquad F=E\wedge dt +B, \quad \label{JFG}\\[4mm] 
F=dA,\qquad \delta F= \mu_\sigma J_\sigma, \qquad \square A=-\mu_\sigma J_\sigma \label{FA}
\end{gather}
where $\sigma=1$ for the Maxwell-Amp\`ere equations, and 0 for the Maxwell-Heaviside equations,
and $\mu_\sigma=\mu_g$ for $\sigma=0$ and $\mu$, the magnetic permittivity, for $\sigma=1$.
\end{theorem}

\medskip

The action integral in terms of the Minkowski metric is given in Landau and Lifshitz, \S27 for the Maxwell-Amp\`ere equations, and in Misner {\it et.al.} Exercise 7.2, for both cases. The action can also be written in terms of the Hodge duality on $\mathbb M^4$.  On $\mathbb M^4$  the volume element is $dv\wedge dx^4=ic \, dv\wedge dt$, so to get a real symmetric inner product on real $p$-forms we take
\begin{equation}\label{Hodgeduality}
(\xi,\eta)=\frac1{ic}\iiiint\limits_{\mathcal K} \xi\wedge \ast\eta, \qquad \xi,\eta\in \Lambda_p( {\mathcal K}), \ \ {\mathcal K}\subset \mathbb M^4.
\end{equation}
Since Maxwell's equations form a hyperbolic system,
we take  the region of integration ${\mathcal K}$ to be the backward ray cone from a point in $\mathbb M^4$. The action for Maxwell's equations is then
\begin{equation}
S=ic\left[\frac12 (F,F)+\,\mu\,(A,J)\right] =\iiiint\limits_{\mathcal K} \frac12 F\wedge\ast F +\mu\, A\wedge \ast J .\label{action}
\end{equation}

The action is imaginary,  and its critical points are obtained via the Principle of Stationary Phase. Denoting the variation of $S$ by $\dot S$, we get
$\dot S=ic[(\dot F,F) +\,\mu\,(\dot A,J)],$
where $(\dot F,F)=(d\dot A,F)=(\dot A,\delta F)$.
Thus $\dot S=ic\, (\dot A, \delta  F+\mu J)$. Letting $\dot A$ vary over all admissible variations, and assumeing these to be a dense set, we get $\delta F+\mu J=0$, which, together with the relation $F=dA$, comprise Maxwell's equations \eqref{FA}. (The third equation in \eqref{FA} is a consequence of the first two.)

The Lagrangian $L$, which is real, is obtained by setting $x^4=ict$ and putting 
$$
L\,dv\wedge dt = \frac1{ic}\left[\frac12 F\wedge\ast F+\mu A\wedge\ast J\right].
$$
In the electrodynamic case, we find
\begin{align*}
A\wedge \ast J=&ice({\bf A}\cdot{\bf p}+\phi\rho)\,dv\wedge dt ;\\[4mm]
F\wedge\ast F=&ic \left({\bf B}\cdot{\bf B}-\frac1{c^2}{\bf E}\cdot{\bf E}\right)\,dv\wedge dt\\[4mm]
=&ic\mu \left(  \mu{\bf H}\cdot{\bf H}-\epsilon{\bf E}\cdot{\bf E}\right)\,dv\wedge dt.
\end{align*}
Hence
\begin{equation}\label{LagrangianElec}
L=\mu \left(  \mu{\bf H}\cdot{\bf H}-\epsilon{\bf E}\cdot{\bf E}\right)+e({\bf A}\cdot{\bf p}+\phi\rho).
\end{equation}

Turning now to the gravitational case, note that $\bar S=-S$ is also an action for Maxwell's equations, and that
\begin{equation}
\bar S=\iiiint\limits_{\mathcal K} \frac12 F\wedge\tilde\ast F +\mu\, A\wedge \tilde\ast J .\label{baraction}
\end{equation}
Thus $\bar S$ is the action for the Maxwell-Heaviside equations \eqref{MHtilde}; and the associated Lagrangian is the negative of that for the electromagnetic field:
\begin{equation}\label{LagrangianGrav}
L=-\mu_g \left(  \mu_g{\bf H}\cdot{\bf H}-\epsilon_g{\bf E}\cdot{\bf E}\right)-({\bf A}\cdot{\bf p}+\phi\rho).
\end{equation}
In particular, the energy of the gravitational field is negative, as noted in \cite{MTW} Exercise 7.2.

 \section{Dark Matter: Red Herring?}\label{galaxies}  The failure of the Schwarzschild metric to model the dynamics of spiral galaxies has been demonstrated dramatically by extensive observations by Rubin et.al. \cite{RubinSC}\cite{Rubin}.

\centerline{\epsfxsize=\linewidth \epsfbox{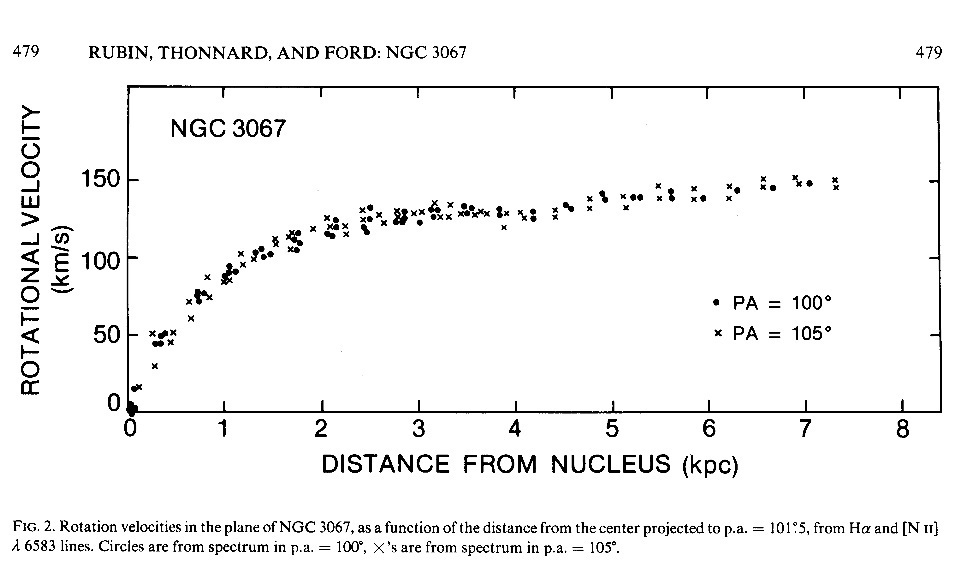}} 
\noindent Reproduced with the permission of the authors and the American Astronomical Society.
\medskip

\noindent  The graph above, called the rotation curve, shows increasing speeds of hydrogen and nitrogen atoms  as measured by shifts in the emission lines $H\alpha$ (hydrogen) and $N\ II$ (nitrogen) going out from the core of the galaxy. The rising curve violates Kepler's third law of planetary motion in a central force field, according to which the velocities should drop off as  $r^{-1/2}$, where $r$ is the mean distance from the central mass. The distance is measured in kiloparsecs, 1 kpc being approximately  3,260 light years. The particle velocities are small compared with the speed of light, but since the gravitational field propagates with the speed of light, it would require 3,260 years for a disturbance to travel 1 kpc. 

The rotation curve, as well as other anomalies in galactic dynamics, have led to a hypothesis of ``dark matter," matter which is not visible -- even yet-to-be discovered forms of matter\footnote{On July 4, 2012 physicists at CERN announced the discovery of a heavy boson with properties similar to the long-sought Higgs boson. Such a discovery is significant for the standard model of elementary particles in quantum physics, but the particle is an artifact of high-energy physics, and decays very rapidly. Such a theory is relevant to the early stages of the universe, when the energy density was very high; but its relevance to galactic dynamics, which is a problem in low energy physics, has yet to be estabished.} which do not interact with the electromagnetic field -- in order to explain the observations. By definition, any ``evidence'' for dark matter in the universe is necessarily indirect, taking the form of dynamics which cannot be explained using current mathematical models. The dark matter hypothesis is sometimes presented as an alternative: 1. Either Newton's Laws of motion are not valid;  or 2. There is additional ``nonluminous'' matter. 

Current research focusses almost entirely on the second proposition, despite the fact that Newton's theory of gravity posits ``action at a distance'',  and cannot hold on a galactic scale. The discovery of some exotic form of ``non-luminous'' matter would thus not resolve the dynamical issues involved. The one conclusion that can definitively be drawn from the data obtained by Rubin and her co-workers, dark matter or not, is that the energy-momentum tensor does not vanish in the outer reaches of galaxies. 

The general theory of the energy-momentum tensor is presented in the monograph by Landau and Lifshitz \cite{LL} \S32,\,\S35, where it is obtained as the Euler-Lagrange derivative of a Lorentz invariant Lagrangian.  Mass does not appear explicitly in $T$, but $T^{00}$ represents energy density. It includes energy of all types: mass, gravitational energy, electromagnetic energy, thermal energy, and  kinetic energy, including rotational kinetic energy.  The general form of the tensor for macroscopic bodies (\S35) is given as the strikingly simple  formula
\begin{equation}\label{Tmacro}
T^{\mu\nu}=(p+\varepsilon)u^\mu u^\nu-pg^{\mu\nu},
\end{equation}
where $\varepsilon$ is the energy density, $p$ is the ``pressure'', and $u^\nu$ is the 4-velocity of the material flow. There are no specifics about the composition of the energy density. 

Equation \eqref{Tmacro} is derived assuming that Pascal's law of hydrostatics is valid in a co-moving frame, that is, a reference frame moving with the material point. In that case, the classical energy-momentum tensor takes the form
$$
T^{\mu\nu}=\begin{pmatrix} \varepsilon & 0&0&0\\ 0 & p& 0&0\\0 & 0& p  & 0\\ 0 & 0 & 0 & p \end{pmatrix}
$$ 
The form \eqref{Tmacro} then follows by a field of Lorentz transformations depending on the point in question. 

Pascal's Law states that the pressure throughout an incompressible fluid at rest is constant. To suggest it holds in the interior of a star or in a turbulent star-forming region such as the Orion nebula is, to say the least, a bit of a stretch. Finally, $T^{\mu\nu}$ is supposed to be the gradient of a Lorentz invariant functional; yet none is given.

The dark matter hypothesis was originally put forward by Zwicky \cite{Zwicky33}.
In \S5 of that paper, Zwicky makes some rough statistical calculations about the Coma cluster in which he compares estimates of its mass obtained by two different methods.  The Coma cluster is comprised of some 800 galaxies at a distance of 45 million light years. Zwicky measures the velocity dispersion of the individual galaxies and, using classical statistical arguments, gets an estimate of the total kinetic energy of the cluster. Assuming the 800 galaxies form a ``gas'' in statistical equilibrium, he uses the classical virial theorem, which implies that the mean kinetic and potential energies of the system are equal, to estimate the total mass of the cluster.  By this means he estimates the average density of the cluster ``to be at least 400 times larger than that derived on the grounds of observations of luminous matter. If this would be confirmed we would get the surprising result that dark matter is present in much greater amount than luminous matter.''

Classical statistical mechanics, based on Newton's theory, plays a central role in Zwicky's analysis;  yet the  Coma cluster is about one million light years across, calling instead for a relativistic statistical mechanics based on the Maxwell-Heaviside equations. Zwicky's model of the Coma cluster as a ``gas'' of point particles raises questions of its own. Galaxies are not point masses; they have internal structure, especially rotational kinetic energy. 

In his 1937 paper  \cite{Zwicky37}, Zwicky returns to the problem of dark matter, listing several caveats. In the abstract he states

\medskip

 ``Present estimates of the masses of nebulae are based on observations of the {\it luminosities} and {\it internal rotations} of nebulae. It is shown that both these methods are unreliable; that from the observed luminosities of extagalactic systems only lower limits for the values of their masses can be obtained, and that from internal rotations alone no determination of the masses of nebulae is possible.'' 
 
 \medskip

He goes on

\medskip

``In order to derive trustworthy values of the masses of nebulae from their absolute luminosities, detailed information on the following three points is necessary.

\begin{enumerate}

\item According to the mass-luminosity relation, the conversion factor from absolute luminosity to mass is different for different types of stars. The same holds true for any kind of luminous matter. In order to determine the conversion factor for a nebula as a whole, we must know in what proportions all the possible luminous components are represented in this nebula.

\item  We must know how much dark matter is incorporated in nebulae in the form of cool and cold stars, macroscopic and microscopic solid bodies, and gases.

\item  Finally, we must know to what extent the apparent luminosity of a given nebula is diminished by the internal absorption of radiation because of the presence of dark matter.''

\end{enumerate}

Zwicky's three points illustrate the complexity inherent in the dark matter mystery, a complexity that is   monotonically increasing as a function of  technology:

\medskip

\begin{quotation} ``In 2005, the Advanced Camera for Surveys instrument of the Hubble Space Telescope finished capturing the most detailed image of the nebula yet taken. The image was taken through 104 orbits of the telescope, capturing over 3,000 stars down to the 23rd magnitude, including infant brown dwarfs $\dots$ A year later, scientists working with the HST announced the first ever masses of a pair of eclipsing binary brown dwarfs. $\dots$ in the Orion Nebula [having] approximate masses of $0.054 M_\odot$ and $0.034 M_\odot$ respectively, with an orbital period of 9.8 days. Surprisingly, the more massive of the two also turned out to be the less luminous.''     Orion Nebula, Wikipedia. \end{quotation}

Rubin expresses support for a dynamical approach to the 
dark  matter mystery: 
\begin{quotation} Currently, the theory of dark matter is the most popular candidate for explaining the galaxy rotation problem. The alternative theory of MOND (Modified Newtonian Dynamics) has little support in the community. Rubin, however, supports the MOND approach, stating ``If I could have my pick, I would like to learn that Newton's laws must be modified in order to correctly describe gravitational interactions at large distances. That's more appealing than a universe filled with a new kind of sub-nuclear particle.'' Vera Rubin, Wikipedia \end{quotation}

The Maxwell-Heaviside equations of gravitation constitute a linear, relativistic correction to Newton's equations of motion; they interpolate between Newton's and Einstein's theories of gravitation, and are therefore a natural mathematical model on which to build a dynamical theory of galactic structures.

\section{Maxwell's Enigma}\label{enigma} In 1905 Poincar\'e \cite{Poincare} observed that the transformations used by Lorentz in his theory of the electron constituted a transformation group. He believed that Lorentz covariance was a fundamental fact of physics; and at the end of his paper, in a section entitled {\it Hypoth\`eses sur la Gravitation}, he proposed a rudimentary  Lorentz-covariant form of the gravitational field that included motion.  He speculated about ``l'onde gravifique, . . .   \'etant suppos\'ee se propager avec la vitesse de la lumi\`ere." ``La force totale", he wrote ``peut se partager en trois composantes, la premi\`ere une vague analogie avec la force m\'ecanique due au champ \'electrique, les deux autres  avec la force m\'ecanique due au champ magn\'etique."  

When Einstein turned to the problem of deriving a relativistic theory of gravity, he introduced the Principle of Equivalence as a fundamental axiom in addition to Lorentz covariance. This states that an observer cannot distinguish between an accelerated reference frame and a gravitational force. By 1907, he had begun to realize that the Principle of Equivalence is incompatible with invariance under Lorentz transformations. Indeed, the Poincar\'e group (the Lorentz group plus space-time translations) is the symmetry group of the Minkowski metric (see  \cite{W}, \S2.1 for a direct proof);  coordinate transformations beyond those in the Poincar\'e group lead to more general metric tensors, and to the curvature of space-time.  

Poincar\'e died in 1912, just as the battle for the Holy Grail of Mathematical Physics of the era, a relativistic theory of gravitation, was beginning to heat up. Minkowski had already passed away in 1909. The banner of special relativity was taken up by three Knights Errant, Gustav Mie, Gunnar Nordst\"om, and Max Abraham in the years 1912-1914.  

There is a lingering perception that gravitation cannot be described by special relativity due to the fact that the energy of the gravitational field is negative (Pais \cite{P}, chapter 13). The issue was prompted by a remark of Maxwell in his paper {\it A Dynamical Theory of the Electromagnetic Field} (\cite{Maxwell}, pp. 570,571). In the section, ``Note on the Attraction of Gravitation," Maxwell compares the energy fields of gravitational attraction and magnetic repulsion of two like magnetic poles. He gives a calculation showing  that  the ``gravitational energy" due to the two bodies is negative, and, though he does not say so explicitly, that it is unbounded below. Maxwell's argument is cryptic from today's perspective, but his conclusion is not:

\begin{quotation} ``The assumption, therefore, that gravitation arises from the action of the surrounding medium in the way pointed out, leads to the conclusion that every part of this medium possesses, when undisturbed, an enormous intrinsic energy, and that the presence of dense bodies influences the medium so as to diminish this energy wherever there is a resultant attraction.

As I am unable to understand in what way a medium can possess such properties, I cannot go any further in this direction in searching for the cause of gravitation.'' \end{quotation}

Abraham \cite{MA}, citing Maxwell's comment, showed that ``Vector Theories" of gravity are inherently unstable. His analysis is couched in terms of partial differential equations, hence more understandable in modern terms. He takes the partial differential equations of a static gravitational field to be
\begin{equation}\label{diffeq_gravity}
{\bf {\rm div}\, F}^g=-\rho, \qquad {\bf F}^g=-\text{grad}\,\phi,
\end{equation}
where $\rho$ is the mass density and $\phi$ the gravitational potential. He then states that in Newton's theory the energy of the resulting gravitational field is 
\begin{equation}
{\mathcal E}=\frac12 \iiint \rho\phi\,dv. \label{interaction}
\end{equation}
(This statement is correct; the proof is similar to that in the electrostatic case \cite{LL}, \S37.) Now equations
\eqref{diffeq_gravity} combine to give Poisson's equation $\Delta \phi =\rho$. Assuming $\rho\in \,L^1(\mathbb E^3)$,
the solution is given by the convolution integral 
$$
\phi(x)=- \int \frac{\rho (y)}{4\pi |x-y|}dy, \qquad x,y\in \mathbb E^3;
$$
and since $\rho\ge 0$, the integral in \eqref{interaction} is negative.
It can be written as
\begin{equation}\label{grav_energy}
{\mathcal E}=-\iint \frac{\rho(x)\rho(y)}{8\pi |x-y|}dxdy, \qquad x,y\in \mathbb E^3.
\end{equation}

The issue raised by Maxwell and Abraham is succinctly illustrated by a simple example:  consider the family of  Gaussians
$$
\rho_\sigma(x)=\left(\frac{\sigma}{\pi}\right)^{3/2} \exp (-\sigma |x|^2), \qquad x\in \mathbb E^3.
$$
These densities have total mass 1, and the energy \eqref{grav_energy} can be computed explicitly
\begin{align*}
{\mathcal E}_\sigma=&-\iiint\frac{dz}{8\pi |z|}\iiint\rho_\sigma(y)\rho_\sigma (y+z)dy\\[4mm]
=&  \iiint\frac{dz}{8\pi |z|}  \left(\frac{\sigma}{2\pi}\right)^{3/2}\exp \left(- \frac{\sigma  |z|^2}2\right)\,dz              =-\frac1{4\pi} \left(\frac{\sigma}{2\pi}\right)^{1/2}.
\end{align*}

In fact, we have

\begin{align*}
\iiint\rho_\sigma(y)&\rho_\sigma (y+z)dy\\[4mm]
 =&\left(\frac{\sigma}{\pi}\right)^3\exp (-\sigma |z|^2)
\iiint \exp (-2\sigma (|y|^2+2 y\cdot z)\, dy\\[4mm]
=&\left(\frac{\sigma}{\pi}\right)^3\exp \left( -\frac{\sigma |z|^2}{2}\right)
\iiint \exp \left(-2\sigma \left |y+ \tfrac{z}2\right |^2\right) dy\\[4mm]
=&\left(\frac{\sigma}{\pi}\right)^3\exp (- \frac{ \sigma |z|^2}2)\left(\int_{-\infty}^\infty \exp (-2\sigma t^2 )\, dt\right)^3
\\[4mm]
=&\left(\frac{\sigma}{2\pi}\right)^{3/2}\exp \left(- \frac{\sigma  |z|^2}2\right).
\end{align*}

This example shows that solutions to Poisson's equation are dynamically unstable: as $\sigma\to\infty$,  the total mass remains the same, but concentrates at the origin, while ${\mathcal E}_\sigma\to -\infty$. Since matter naturally aggregates under the force of gravity, this implies that matter, in the absence of forces other than gravity, collapses to a point.

Pais suggests that the negative energy issue was the stumbling block to a Lorentz invariant field theory of gravity:
\begin{quotation} ``Maxwell's wise words were not generally heeded, not even by physicists of great stature. Oliver Heaviside discussed the gravitational-electromagnetic analogy without mentioning the negative energy difficulty. So, remarkably, did Lorentz in one of his rare speculative papers . . .  As late as  1912, it was still necessary to show that all these vector theories made no sense because of Maxwell's negative energy difficulty.'' \end{quotation}

The negative energy of the gravitational field was cited by Abraham in his dispute with Einstein over Lorentz invariance as a fundamental principle of physics, and had gotten entangled in the fray. But  the negative energy is neither an obstacle to a field theory of gravitation -- classical potential theory is a counterexample -- nor is it resolved by the General Theory. Indeed the issue is central to modern cosmology. It led Einstein to introduce his cosmological constant in order to ``stabilize" the universe. The same instability underlies  the theory of black holes, which finds its genesis in the singularity of the Schwarzschild metric. 

Einstein's adherence to the Principle of Equivalence forced him to abandon the Lorentz group in favor of a more general class of observers; while Abraham, Mie, and Nordstr\"om continued to hold that  Lorentz invariance was a fundamental hypothesis of nature.  (see the discussion in Thorne \cite{Thorne}, p. 115.) But the issue, like the negative energy issue, is not so clear cut.

If the Principle of Equivalence knocks out the Maxwell-Heaviside gravitational theory, it knocks out Maxwell's electromagnetic theory as well -- something no one, not even Einstein himself, would claim.  Strict Lorentz invariance would not allow for force or acceleration in electrodynamics; yet a fundamental deduction from Maxwell's equations is the existence of electromagnetic waves, generated by accelerated charges.

Accelerated reference frames are in fact implicit in Maxwell's equations: here are two arguments. First, Maxwell's equations form a first order hyperbolic system. If the first order system is reduced to a linear second order equation, as Maxwell did, one finds that it contains second order time derivatives, a.k.a. acceleration. Second, the time independent Maxwell equations describe the static  fields and are themselves invariant under Lorentz transformations.  All inertial observers see a static field, with only the relative strength, position and  direction of the electric and magnetic fields changed. Thus, the dynamic and static Maxwell's equations are each invariant under Lorentz transformations, so to go from one to another requires a transformation to an accelerated reference frame. 

The second proof is motivated by a ``Machian''  {\it Gedankenexperiment}. Consider a binary star, rotating in the $x-y$ plane, with the center of gravity of the two stars at the origin. Consider two observers on the $z$-axis, one whose coordinate system is fixed to the $x,y,z$ axes; the other whose coordinate system is tied to the circling stars. 
The first sees the two stars rotating about each other; the second sees them sitting motionlessly, suspended in space; but to go from one to the other requires a transformation to an accelerated reference system.

The atmosphere, moving with the Earth, is in an accelerated reference frame and experiences a force, the Coriolis force. The Corolis force in geophysics is thus a familiar example of Einstein's  Principle of Equivalence in Newtonian mechanics.  In short, the entire subject of dynamics, inasmuch as it  involves force and acceleration, is incompatible with inertial reference frames. Accelerated reference frames are indeed synonymous with the curvature of space-time, but by Einstein's Equivalence Principle, space-time is virtually flat if the acceleration is not too great, and linear theories suffice.

Finally, Einstein's computation of the precession of the perihelion of Mercury was a signature achievement.  It validated the General Theory and demonstrated the fundamental nonlinear character of the gravitational field; but it does not invalidate the Maxwell-Heaviside theory in weak fields. Thorne (\cite{Thorne}, p. 95) points out that the total advance of the perihelion is 1.38 seconds of arc per revolution, 1.28 of which can be accounted for by Newtonian theory.  So Newton's theory is in fact remarkably good!  No one is suggesting that Newton's linear, non-relativistic theory be abandoned because of this one small lapse; so why should the Maxwell-Heaviside theory, taken as a linear approximation in a weak field, be rejected on these same grounds? 

Abraham's paper contains a comprehensive bibliography of the literature devoted to vector models  of gravitation. ``This paradoxical conclusion [the negative energy of the field] is characteristic of the vector theory of the gravitational field,'' writes Abraham. He goes on to discuss the gravitational theories of Lorentz, Poincar\'e and Minkowski in some detail, concluding that ``the laws of interaction of attracting masses drawn up by these pioneers of relativity are accordingly very well reconciled with the vector theory of gravitation sketched above."  
 
\section{Acknowledgements} I would like to express my thanks to the Editorial Board for their invitation to submit this paper to the memorial volume for Klaus Kirchg\"assner. I first visited Klaus in 1970 when he was at the Ruhr-Universit\"at in Bochum. We became instant friends, and we enjoyed many discussions over the years, not only about mathematics, but about history, music, German Romanticism, people (Kepler, Einstein, Newton, Mozart, etc.) Klaus was a prolific mathematician, with high intellectual standards, and his work, especially his contributions to the theory of nonlinear waves, has set the standard.  A mentor, a gracious host, an organizer,  a promoter of science and mathematics, he was always interested in what everyone else was doing. He had a wonderful sense of humor. He will always be with us.

Everyone associated with Klaus shared his interest in the applications of mathematics to physical problems. All of us have worked in some area or another of dynamical systems or continuum mechanics, with analysis and differential equations as a central tool. The techniques we acquired are directly applicable to problems in modern Cosmology.  

 I thank a number of people for useful comments and stimulating conversations which have been incorporated into this article:  Jacek Szmigielski, Adrian Korpel, Jim Wheeler, Paul Faris, Joel Smoller, and Jeff Rauch.

  \bibliographystyle{plain}
\bibliography{Maxwell}

\end{document}